\documentclass{llncs}
\usepackage{llncsdoc}
\usepackage{graphicx}
\usepackage[export]{adjustbox}
\usepackage{setspace}
\usepackage{etoolbox}
\begin{document}
%\markboth{\LaTeXe{} Class for Lecture Notes in Computer
%Science}{\LaTeXe{} Class for Lecture Notes in Computer Science}
\thispagestyle{empty}

\newcounter{save}\setcounter{save}{\value{section}}
{\def\addtocontents#1#2{}%
\def\addcontentsline#1#2#3{}%
\def\markboth#1#2{}%
\title{Feature Model-to-Ontology for SPL Application Realisation}

%\author{Ivar Ekeland\inst{1} \and Roger Temam\inst{2}}
\author{Iis Afriyanti, Faiq Miftakhul Falakh, Ade Azurat, Bravyto Takwa}

\institute{Faculty of Computer Science, Universitas Indonesia\\Depok, Indonesia\\
\email{iisafriyanti@ui.ac.id, faiqmiftakhul@cs.ui.ac.id, ade@cs.ui.ac.id, bravyto.takwa@ui.ac.id}}

\maketitle
\begin{abstract}
Feature model are widely used to capture commonalities and variabilities of artefacts in Software Product Line (SPL). Several studies have discussed the formal representation of feature diagram using ontologies with different styles of mapping. However, they still focused on the ontology approach for problem space and keep the solution space aside. In this paper, we present the modelling of feature model using OWL ontology and produce an application based on the ontology. Firstly, we map the features in a running example feature diagram to OWL classes and properties. Secondly, we verify the consistency of the OWL ontology by using reasoning engines. Finally, we use the ontology as an input of Zotonic framework for application realisation.
\end{abstract}
\section{Introduction}
Software Product Line (SPL) engineering is \textit{"a paradigm to develop software applications (software intensive-systems and software products) using platforms and mass customisation} \cite{pohl:SPLE}. SPL also can be considered as a representation of all applications in a particular domain that can be personalised for a particular need \cite{czarnecki}. This paradigm promises shorter time with lower cost, yet higher quality compared with the individual software development \cite{pohl:SPLE}; that is generally structured from requirement gathering to deployment phase for each individual software. 

According to Pohl, \cite{pohl:SPLE} the paradigm of SPL runs in two main development processes, namely \textit{domain engineering} and \textit{application engineering} . In the domain engineering that can be called as \textit{problem space}, the development focuses on establishing the features or characterise the applications by identifying commonalities and variabilities artefacts for a certain domain. It captures the common artefacts that can be reused for all applications and manage flexibility of product that may fits for different stakeholders but still in a targeted domain. On the other hand, the application engineering or so-called \textit{solution space} is to realise the features that is previously identified in the domain engineering into an application.

Aforementioned, capturing commonalities and variabilities is a key point in SPL and it can be visualised in a feature diagram; a tree-like model with nodes illustrates features in a product line. A feature diagram also stores some constraints and rules to represent a valid application. For example, the constraint representation for a mandatory and optional features for a product. However, one of challenges of feature diagram is lack of formal semantic \cite{verifying}\cite{applyingSW}, thus cannot be understood by machine. %Since then, several studies adopt the ontology to put the meaning into the feature models.

Ontology, defined as \textit{"an explicit specification of a conceptualisation"}\cite{Gruber1993}, has a considerable attention to be investigated its opportunity for formal representation of feature diagram. In \cite{verifying}, the feature diagram is transformed into an OWL ontology to verify the consistency of feature diagram by using reasoning tools in Prot\'eg\'e; a tool to implement OWL ontology and check its consistency. Similar transformation was conducted by \cite{applyingSW} that added Semantic Web Rule Language (SWRL) rules in OWL feature model to infer some interesting facts related with the designed features. The study of ontology for advancing feature diagram grows even more; not only for checking the consistency and inference, but also for sharing knowledge base of features in a certain domain \cite{safmdl}\cite{enrichment}  that can be investigated its practicability for features integration between two or more domains.

Most of current works focus on feature diagram-to-ontology mapping for formal representation of the problem space and keep the solution space aside. Nevertheless, in \cite{viewsontologies} analysed that the mapping mechanism can be undertook for any types of purposes, including mapping to ontologies that is closer to the solution space. Therefore, in this study we attempt to map the feature models to ontology and carry out the feature into a product as part of application realisation in SPL framework.

In this study, we employ existing AISCO feature diagram as a running example followed by mapping it into an OWL-ontology. Thereafter, we translate the ontology into an application by utilising Zotonic\footnote{http://docs.zotonic.com/en/latest/index.html}, a Content Management System with pragmatic semantic web as its data model.

The rest of this paper is structured as follows, in Section 2 explains current studies on feature model-to-ontology mapping. Section 3 presents AISCO feature model to represent features in charity organisation domain as part of our running example. In Section 4, we provide a one-to-one mapping from the feature model to the OWL-based ontology. In Section 5, we explain the design of our system. Finally, Section 6 is our concluding remark.

\section{Feature Diagram to Ontology Mapping Studies}
Feature diagram and ontology are in the same level of abstraction that provide meta-information of knowledge in a certain domain \cite{viewsontologies}. Feature diagram is a tree-like visualisation of feature model used to define the commonalities and variabilities of features in a product line. The nodes represent features and the edges portrays the relation constraint between features (see Table \ref{FD_constraints}). On the other hand, ontology represents the knowledge for a certain domain in a network-like model. It consists of the relationship between concepts that can be illustrated in a triple statement \textit{subject, predicate, and object}. However, ontology can be formally presented in Web Ontology Language (OWL)\footnote{https://www.w3.org/TR/2012/REC-owl2-primer-20121211/}; an ontology language with Description Logics as its foundation. \textit{Subject} of statement belongs to \textit{class} in OWL, while the predicate may be called \textit{object property} if the subject relates with another class membership and \textit{datatype property} if the subject relates with data values. Within OWL ontology, it can be extended with some rules assigned (e.g with Semantic Web Rule Language\footnote{https://www.w3.org/Submission/SWRL/}) that can be executed to infer more knowledge (see Table \ref{DL_axiom}).

\begin{table}[h]
	\centering
	\caption{Relations Constraints in Feature Diagram}
	\label{FD_constraints}
	\begin{tabular}{ p{2cm} p{6cm} p{2cm} }
		\hline
		Relation & Explanation & Notation \\ \hline
		\vspace{0pt} Mandatory &  \vspace{0pt} The feature \textit{F} must be added in a product line &  \includegraphics[width=1cm,valign=T]{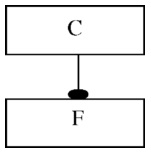} \\
		\vspace{0pt} Optional & \vspace{0pt} The feature \textit{F}  may or may not be added in a product line & \vspace{0pt} \includegraphics[width=1cm,valign=T]{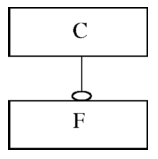} \\ 
		\vspace{0pt} Or & \vspace{0pt} One or more features can be selected in a product line & \includegraphics[width=1cm,valign=T]{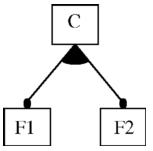} \\
		\vspace{0pt}Alternative & \vspace{0pt}One and only one feature from a set of \textit{alternative} can be included if their parents feature is included in a configuration & \includegraphics[width=1cm,valign=T]{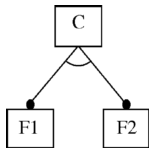} \\ 
		%Include & ... & ... \\ 
		\vspace{0pt}Requires & \vspace{0pt} The presence of feature \textit{F1} requires feature \textit{F2}, thus the feature \textit{F1} will not presented without feature \textit{F2} &  \includegraphics[width=1cm,valign=T]{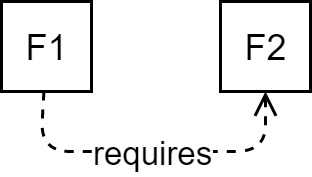}\\
		\vspace{0pt}Excludes & \vspace{0pt} To have feature \textit{F1} to be presented, the product line must exclude feature \textit{F2} &  \includegraphics[width=1cm,valign=T]{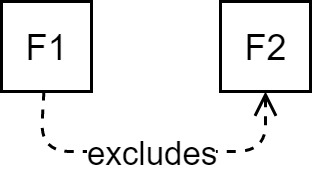}\\ \hline
	\end{tabular}
\end{table}

The study of \cite{viewsontologies} investigated that the relationship between feature model and ontology are closely related. The study showed the mapping on notational spectrum of feature diagram to ontology. Moreover, they illustrated that considering the expressiveness of knowledge representation, ontology is richer than feature diagram; the feature diagram constraint are limited for the features in a product line, but the ontology is for any domain.  They argued that the mapping from feature diagram to ontology is attainable for all types of mapping, but the mapping might be unmanageable if the ontology is closer to implementation.

Meanwhile, \cite{verifying} attempted to use ontology to verify the consistency of feature models since manual verification in feature model would be difficult. They translated the feature model into OWL to verify the model using a reasoner (e.g Pellet reasoner that already plugged in Prot\'eg\'e). Since then, the focus of these kind of works mainly on global ontology of feature model to store the knowledge base of features models. In \cite{applyingSW} applies ontology of feature model to analyse the feasibility of distributed features, thus they designed a shared feature model ontology that can be reused. Similarly, \cite{safmdl} introduced Semantic Annotations for Feature modelling Description Language (SAFMDL) to provide semantic for feature model. In addition, \cite{enrichment} stressed the enrichment of features diagram that can be brought from external domain.

The studies presented thus far provide two mapping styles: OWL classes and OWL individuals style. In OWL classes-based, each feature of a feature model is mapped into an OWL class \cite{verifying}. For example, $ProgramData$ feature in Fig. \ref{fig_FDAISCO} is mapped into \textit{ProgramData} class in OWL ontology. Meanwhile, OWL individuals-based considers all features as individuals or instances of \textit{Feature} class. Such mechanism is applied by \cite{applyingSW},\cite{safmdl}, and \cite{enrichment}.

The mapping decision remains unclear that no strong reasons leading the design. It might be the purpose of their works such as to see reasoning support over feature classes that prefers to adopt OWL classes mapping \cite{verifying}. On the other hand, OWL individuals is more suitable for dynamic features that might change frequently. In this studies, however, we adopt OWL class mapping to translate feature classes into realisation with AISCO feature diagram as a running example.

\section{\textbf{Running Example: AISCO Product Line Feature Model}} AISCO (Adaptive Information System for Charity Organization) is a software system that helps charity organisations to publish their activities and generate financial reports. The organisations have shared characteristics, yet their activities are widely varied. In this paper, we employ the AISCO as a running example to demonstrate the modelling process from the feature diagram to OWL ontologies.

In AISCO product line, the feature for monitoring a charity program (\textit{ProgramData}) is \textit{mandatory} since generally all organisations need this feature. The organisations may have programs with different scheduling behaviour, which could be \textit{periodic} (weekly, monthly, or yearly), \textit{eventual} (a program that can be run any time), or \textit{continuous} (a program that runs gradually). All of them require different system behaviours. Moreover, it is possible that an organisation runs several programs for different natures; there is an organisation that only has periodical program, another organisation has eventual program, and another type of organisation may have these three kinds of program. They modelled the different scheduling types as \textit{optional} child features of $ProgramData$. Mandatory feature is also applied for a publication system ($PublicationSystem$) and financial reporting ($FinancialReport$) since these are required by all organisations. Other optional features include objective/target data that specifies the recipient’s information, and donation reporting. They also modelled notification feature for the member of organisation ($MemberNotification$) that requires $Donor$ and $AutomaticReport$ feature requires donation summary ($Summary$) summary from the system. Based on the above feature classification, a feature diagram for the AISCO applications can be defined as shown in  Fig. \ref{fig_FDAISCO}. The feature diagram is based on the widely-used technique FODA that originally presented by Kang et al \cite{kang}.

\begin{figure}
\makebox[\textwidth][c]{
\includegraphics[width=16cm, height=7cm] {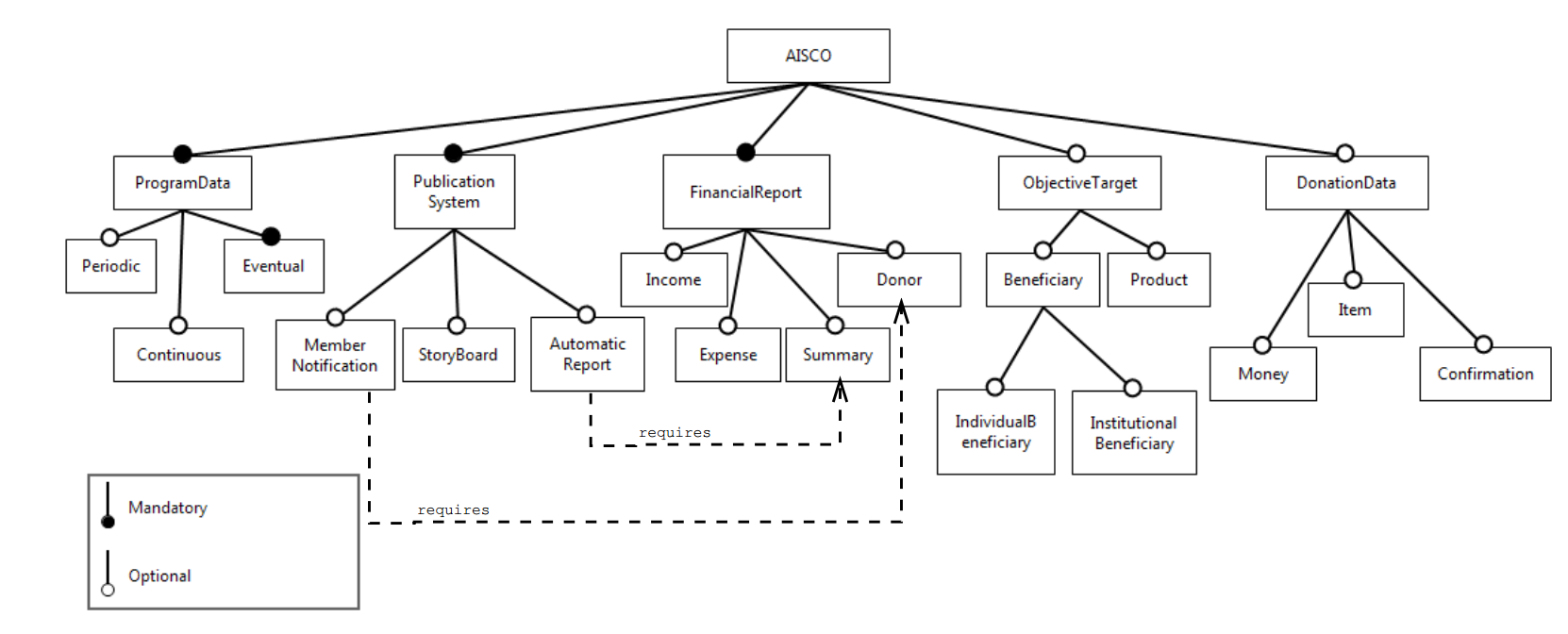}
}
\caption{Feature Model for the AISCO Running Example}
\label{fig_FDAISCO}
\end{figure}

The feature diagram is represented in graphical tree-like notations that shows the hierarchical organisation of features. The root of the tree represents a concept node. All other nodes represent different types of features.

\section{\textbf{Feature diagram modelling to OWL ontology}}
%OWL is one of the most expressive languages for specifying, publishing, and sharing ontologies \cite{tenorio}. An OWL ontology consists of classes, properties, and individual. The classes represent sets of objects or individuals in the domain of discourse. The properties are used to relate between objects. The properties in OWL consists of two kinds properties that have different functions. The first one is object property to link between two objects that member of classes. The second one is datatype property two connect between a member of class and literal data. 

%The properties are binary relations that link individuals, which are subsets of the cross product of the objects in the domain of discourse.

In this section, we describe the transformation process from a feature diagram into OWL ontology classes and properties. First, the nodes in feature diagram are identified and modelled as OWL classes. Each class is defined as mutually disjoint since we assume that features with different names are distinct. Second, for each node, we create a Rule class associated with the nodes. Third, for each of the edges in the feature diagram, we create object property. The range of the property is the respective feature class. 

All axioms of the ontologies described in this paper are formally defined in Description Logics (DL). The OWL syntax used to represent such axioms is summarised in Table \ref{DL_axiom}. In general, for a parent feature A and its child features $B_1, …, B_n$, the initial transformation produces the ontology model as shown in Table \ref{init_ontology}.

\begin{table}
	\centering
	\caption{Summary of Description Logic axioms syntax for ontology}
	\label{DL_axiom}
	\begin{tabular}{ p{2cm} p{9cm} }
		\hline
		Notation & Explanation        \\ \hline
		$\top$ & Superclass of all OWL classes \\ $A \sqsubseteq B$ & A is a subclass of B \\ $A \sqsubseteq \neg B$ & A and B are disjoint classes \\ $A \sqcap B$ & Class intersection \\$A \sqcup B$ & Class union \\$A \equiv B$ & Class equivalence \\ $\top \sqsubseteq \forall P.A$ & Range of property is class A\\ $\exists / \forall P.A$ & someValuesFrom/allValuesFrom restriction, giving the class that for every instance of this class that has instances of property P, some/all of the values of the property are members of the class A \\ \hline
	\end{tabular}
\end{table}

\begin{table}[h]
	\small
	\centering
	\caption{Initial ontology modelling from feature diagram}
	\label{init_ontology}
	\makebox[\textwidth][c]{
	\begin{tabular}{p{2cm}|p{2cm}|p{4cm}|p{7cm}}
		\hline 
		Description & Model from first and second step   & Model from third step     &  Example AISCO ontology model  \\ \hline\hline
		Initial model for parent feature A &   \parbox[t]{2cm}{$A \sqsubseteq \top$ \\ $ARule \sqsubseteq \top$}    &  \parbox[t]{4cm}{$hasA \sqsubseteq ObjectProperty$ \\ $\top \sqsubseteq hasA.A$ \\ $ARule \equiv \exists hasA.A$}     &   \parbox[t]{7cm}{$AISCO \sqsubseteq \top$ \\ $AISCORule \sqsubseteq T$ \\ $hasAISCO \sqsubseteq ObjectProperty$ \\ $T \sqsubseteq hasAISCO.AISCO$ \\ $AISCORule \equiv \exists hasAISCO.AISCO$}  \vspace{3px} \\ \hline
		Initial model for child feature $B_i, ..., B_n$& \parbox[t]{2cm}{$B_i \sqsubseteq \top$ \\ $B_iRule \sqsubseteq \top$}      & \parbox[t]{4cm}{$hasB_i \sqsubseteq ObjectProperty$ \\ $\top \sqsubseteq hasB_i.B_i$ \\ $B_iRule \equiv \exists hasB_i.B_i$ }       &  \parbox[t]{7cm}{$ProgramData \sqsubseteq T$ \\ $ProgramDataRule \sqsubseteq \top$ \\ $hasProgramData \sqsubseteq ObjectProperty
		$ \\ $\top \sqsubseteq hasProgramData.ProgramData$ \\ \vspace{-15px}\begin{tabbing}$Prog$\=$ramDataRule \equiv$\\ \>$\exists hasProgramData.ProgramData$\end{tabbing}}  \\ \hline
		Initial model for ensure disjointness& \multicolumn{3}{l}{\parbox[t]{10cm}{$A \sqsubseteq \neg B_i, for 1 \leq  i \leq n$ \\ $B_i \sqsubseteq \neg B_j, for 1\leq i, j\leq n\wedge i\neq j$}} \\ \hline
	\end{tabular}}
\end{table}

There are three types of feature relations in AISCO feature diagram that we are going to transform into an ontology: mandatory, optional, and requires. In following, the feature relations using the ontology are defined. The general definition of each of the four feature relations are shown based on the above feature ontology.

\textbf{Mandatory}. For each of the mandatory features $B_1, ..., B_n$ of a parent feature A, one N constraint in ARule is used to model it. It is a \verb|someValuesFrom| restriction on $hasB_i$, stating that each instance of the rule class must have some instance of $B_i$ class for $hasB_i$. The following ontology fragment shows the modelling of mandatory feature set and parent feature A.\\
{\small $ARule \sqsubseteq \exists hasB_i.B_i, for 1 \leq i \leq n$} 
\\

Fig. \ref{fig_FDAISCO} shows that root node $AISCO$ has a mandatory child $ProgramData$, which is a non-leaf node. We create additional class for these non-leaf nodes.\\
{\small $ProgramData \sqsubseteq \top$ \\ 
$ProgramDataRule \sqsubseteq \top$\\
$hasProgramData \sqsubseteq ObjectProperty$\\
$\top \sqsubseteq hasProgramData.ProgramData$\\
$ProgramDataRule \equiv \exists hasProgramData.ProgramData$\\
$AISCORule \sqsubseteq \exists hasProgramData.ProgramData$ }\\

\textbf{Optional}. According to the feature modelling of \cite{verifying}, for each of the optional features $B_1, ..., B_n$ of a parent feature A, no additional statements are required to model this relationship.\\
{\small $Bi \sqsubseteq \top $\\
$B_iRule \sqsubseteq \top$\\
$hasB_i \sqsubseteq ObjectProperty$
$B_iRule \equiv \exists hasB_i.B_i,for 1\leq i \leq n$}
\\

It can be seen from Fig. \ref{fig_FDAISCO} that feature $DonationData$ is an optional feature for $AISCO$. Thus, $DonationData$ may or may not be included in a product line configuration of AISCO. As usual, we also create class $DonationDataRule$ for $DonationData$, but no new restriction on AISCO.\\
{\small $DonationData \sqsubseteq \top$\\
$DonationDataRule \sqsubseteq \top$\\
$hasDonationData \sqsubseteq ObjectProperty$\\
$DonationDataRule \equiv \exists hasDonationData.DonationData$}\\

{\textbf{Requires}. For a given feature $A$ and a set of features $B_1;... ; B_n$ that $A$ requires, besides the necessary and sufficient condition that binds $ARule$ to $A$, it is certain that each of the $B_i$ features appears in a configuration if $A$ is present. In Fig. \ref{fig_FDAISCO}, the feature $MemberNotification$ requires $Donor$. Its OWL representation is as follow.}

{\small
\begin{tabbing}
$ARule \sqsubseteq \exists hasBi.Bi, for 1<=i<=n$\\\\
$MemberNotificationRule \sqsubseteq \top$\\
$Mem$\=$berNotificationRule \equiv \exists hasMemberNotification.$\\ 
\> $MemberNotification$\\
$MemberNotification \sqsubseteq \exists hasDonor.Donor$\\
\end{tabbing}
}
\vspace{-12px}
{We evaluate the ontology and enrich it with additional object property to accommodate the needs of the application with respect to charity organisation domain. The OWL ontology is developed using Prot\'eg\'e ontology editor and result an OWL file. There are also some additional datatype properties that belong to some classes. One of the datatype properties is \verb|total|. \verb|Total| would be a trigger for summation function in the Zotonic composer system. We use Pellet Reasoner\footnote{https://github.com/stardog-union/pellet} for checking the consistency of ontology. For the final encoded ontology, Pellet checks the consistency and show that encoded definitions are consistent. Later, we use the OWL as an input for the Composer system, so that the created Zotonic site will have components which associated with the ontology design.}

\section{\textbf{System Design}}
Now we have the ontology model derived from the transformation in the previous section, we move forward to explain the design of the system that utilise the ontology as an input. We describe the system design into four distinguish parts: an overall system design, a design for mapping OWL elements to Zotonic elements, a simple business logic design, and a design for additional script in the installation process of Zotonic.

\begin{figure}
\centering
\includegraphics[width=3in]{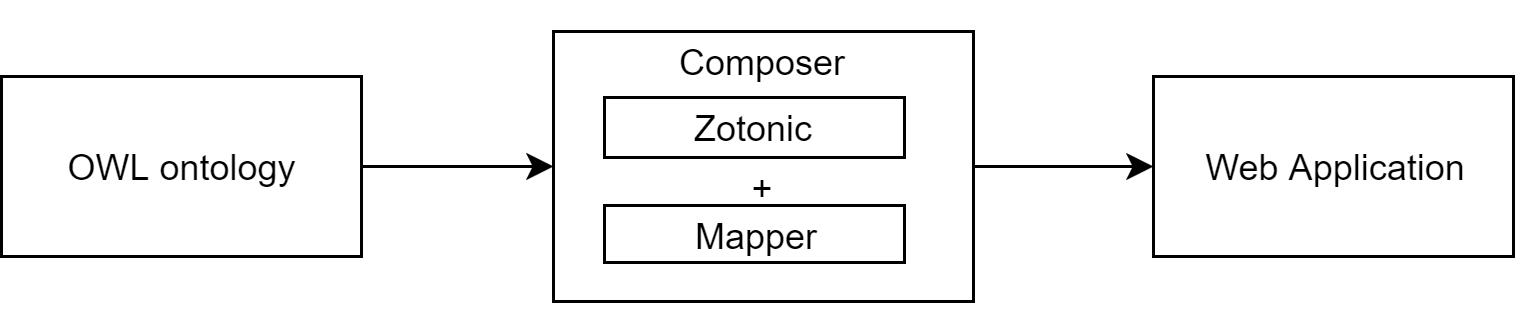}
\caption{Overview of System Design}
\label{systemdesign}
\end{figure}

The main component of the system is composer, which is the Zotonic system and an additional script called \textit{mapper}\footnote{gitlab.com/andrikurniawan/adaptor-ontology-to-webservices}. Within the mapper in the composer, the OWL ontology is automatically transformed into web application. The output of the system is a web application that has similar structure with the input ontology.

\textbf{OWL-to-Zotonic elements mapping}. Mapper is a component to map the elements in OWL ontology to elements in Zotonic framework. The mapped elements are elements in OWL which could build a web application structure. Instances or the contents of the classes might be filled later by the user.

\begin{figure}
\centering
\includegraphics[width=3in]{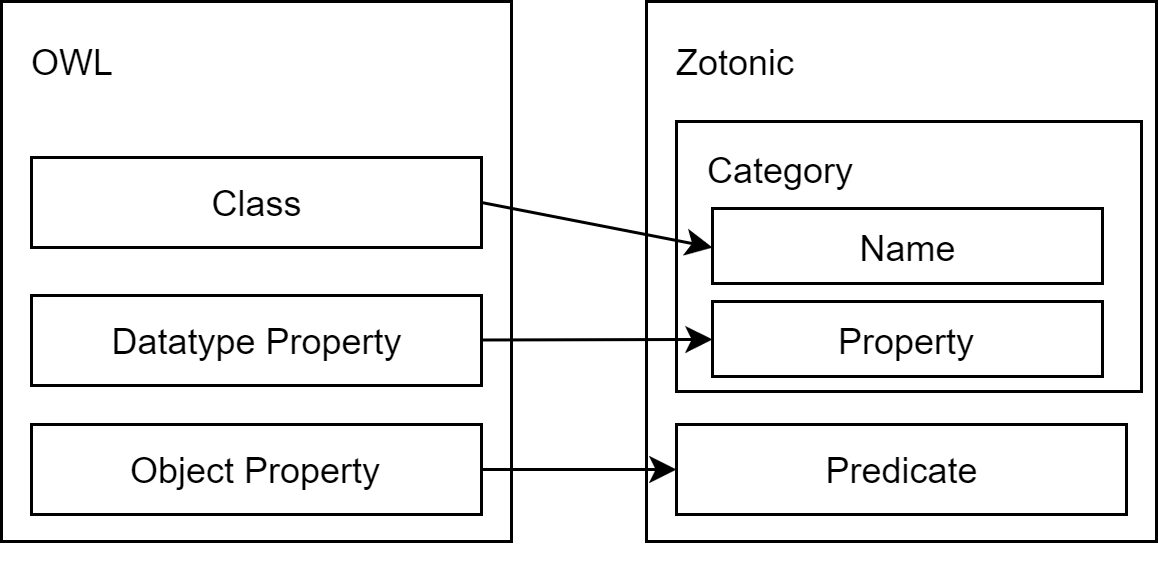}
\caption{Mapping from OWL to Zotonic elements}
\label{owlzotonicmapping}
\end{figure}

Thereafter, the OWL class will be a model of an object in ontology. A class may also have instances as a part of itself. Therefore, class will be an important element which is mapped to Zotonic element. Class will be mapped as a Category name. In Zotonic system, each Category will have instances that represent pages on the web. The pages are easily created by the user.

Zotonic Category has defined properties, yet they are limited to common properties that related with a blog page. To add a particular property in a category, user has to add manually in a Zotonic admin view. To automate this process, datatype property in OWL will be mapped to Category property. As a result, generated web application has complete properties as defined in input OWL ontology.

Default Category properties are shown in Fig. \ref{owlzotonicmapping}.  If the properties have values, then the values will be shown. In this page, datatype properties in input OWL will be augmented as a data form field to fill the values of Category properties.

The last mapped element is object property. In OWL, object property shows relationship between classes. Object property defines a rule that constrain a class as domain and another class as range. The function is similar with predicate function in Zotonic. In Zotonic, predicate represents relationship among instances of Category. Instance, which is a page in Zotonic, connects with another instance using predicate. However, by the Zotonic rule, not all page categories can be connected with others. Therefore, object property in OWL will be mapped to predicate in Zotonic. The result of the object property mapping will be added on the list of predicate in Zotonic.

\textbf{Business Logic Design}. To bring automation advantage for the user, it is required to build simple business logic to facilitate basic arithmetic. One way to build business logic is by detecting similar name between datatype properties and predefined business logic as a trigger. Datatype properties are selected to be business logic based on their names. Then mapper will manage the selected datatype properties as it manages other datatype properties. The difference will be shown on the Zotonic admin page. The business logic will be a field that has been filled automatically by some value. The value is the result from running the business logic that match with particular datatype property.

\textbf{Additional Script}. To generate a web based application from Zotonic automatically, system has to run a build script and create site script as shown in Fig. \ref{systemdesign}. First step is running the build script, and the second step is to run create site script. The script will compile documents in Zotonic initial framework. The initial framework will build the web application structure. By calling a command in terminal, the new site can be accessed on our web browser. 

\begin{figure}
	\centering
	\includegraphics[width=3in]{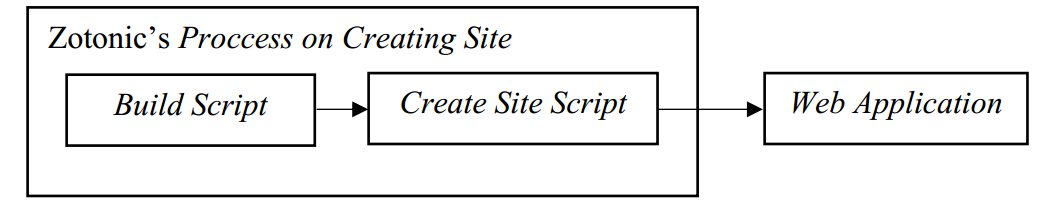}
	\caption{Zotonic Process in Creating a Site}
	\label{zotonicprocess}
\end{figure}

Mapper has to be separated because mapping process from OWL elements could not be directly managed in one call. Therefore, mapper has to be run before compile process and after creating site as viewed in Fig. \ref{mapper}. Both mapper will execute mapping process from same OWL document. The first mapper named Class and Object Property Mapper. It maps every category, predicate, and predicate rule from OWL to Zotonic initial framework in the document z\_install\_data.

\begin{figure}
\centering
\includegraphics[width=3in]{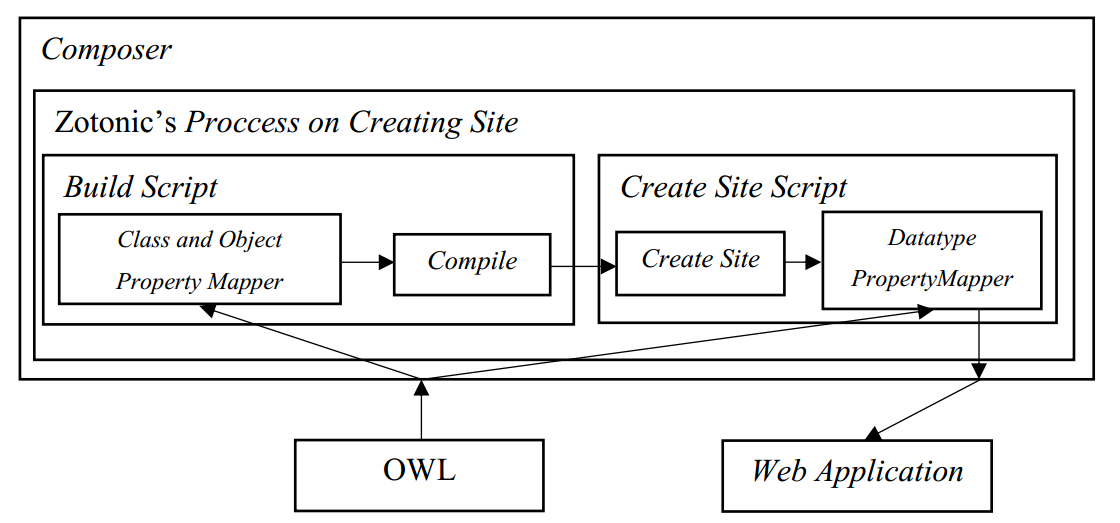}
\caption{Zotonic Process in Creating a Site}
\label{mapper}
\end{figure}

The second mapper called Datatype Property Mapper. It maps datatype property in OWL to admin or user view template page. The mapping process encompasses business logic build from datatype properties. The second mapper has to be run on create site script right after the site has been built. By the help of both mapper, user is only required to install Zotonic based on predefined steps. The result will be a web application that has similar structure with input ontology. Fig. \ref{categoryaisco} and \ref{predicateaisco} shows that every class and object property represented in the OWL ontology is successfully mapped into category and predicate in the produced web application.

\begin{figure}
	\includegraphics[width=12cm, height=7cm]{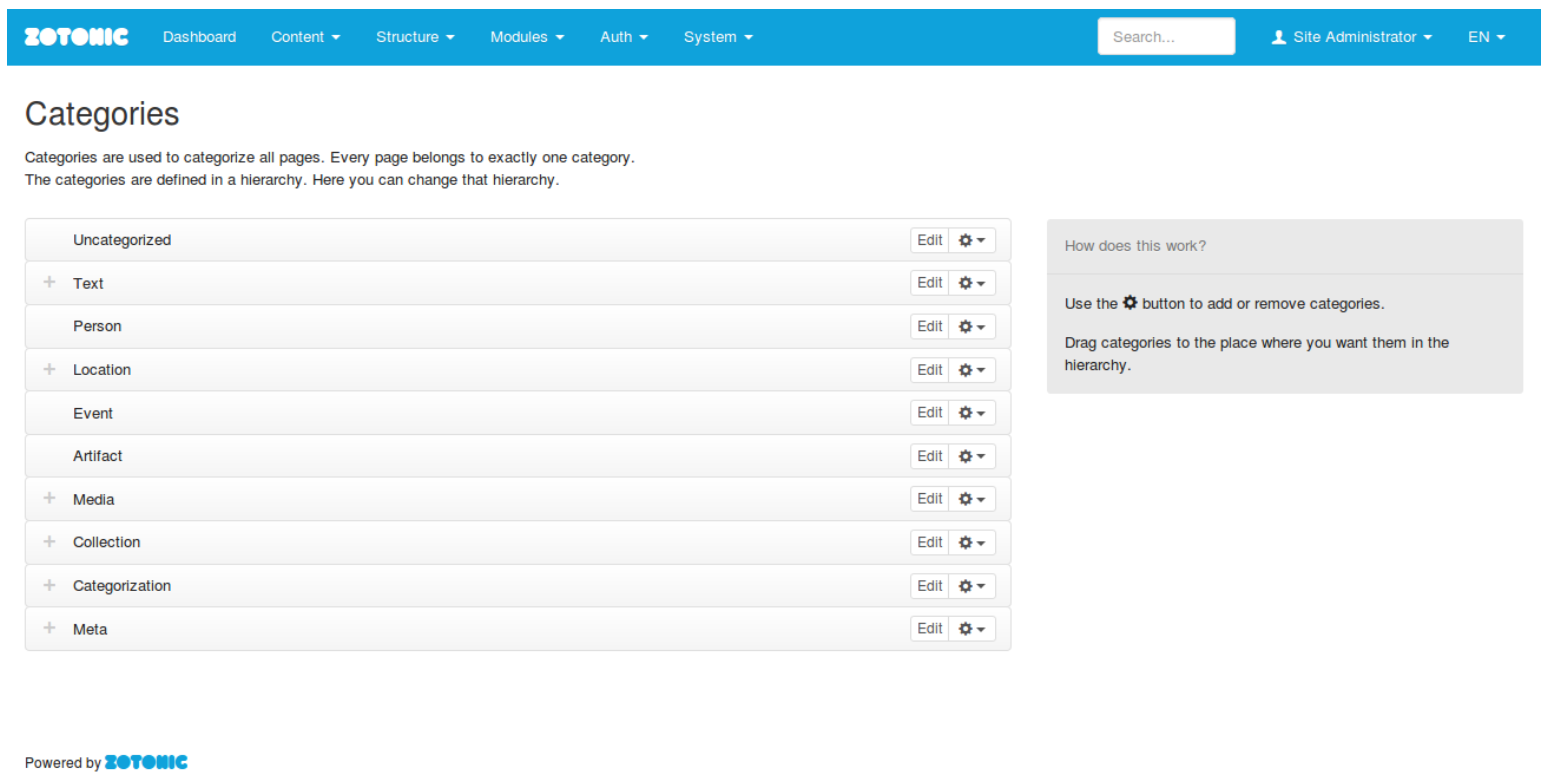}
	\caption{Screenshot of Category list in Zotonic page}
	\label{categoryaisco}
\end{figure}

%\begin{figure}
%	\hspace{-1cm}
%	\includegraphics[width=14cm, height=7cm] {FDAISCO}
%	\caption{Feature Model for the AISCO Running Example}
%	\label{fig_FDAISCO}
%\end{figure}

\begin{figure}
	\includegraphics[width=12cm, height=7cm]{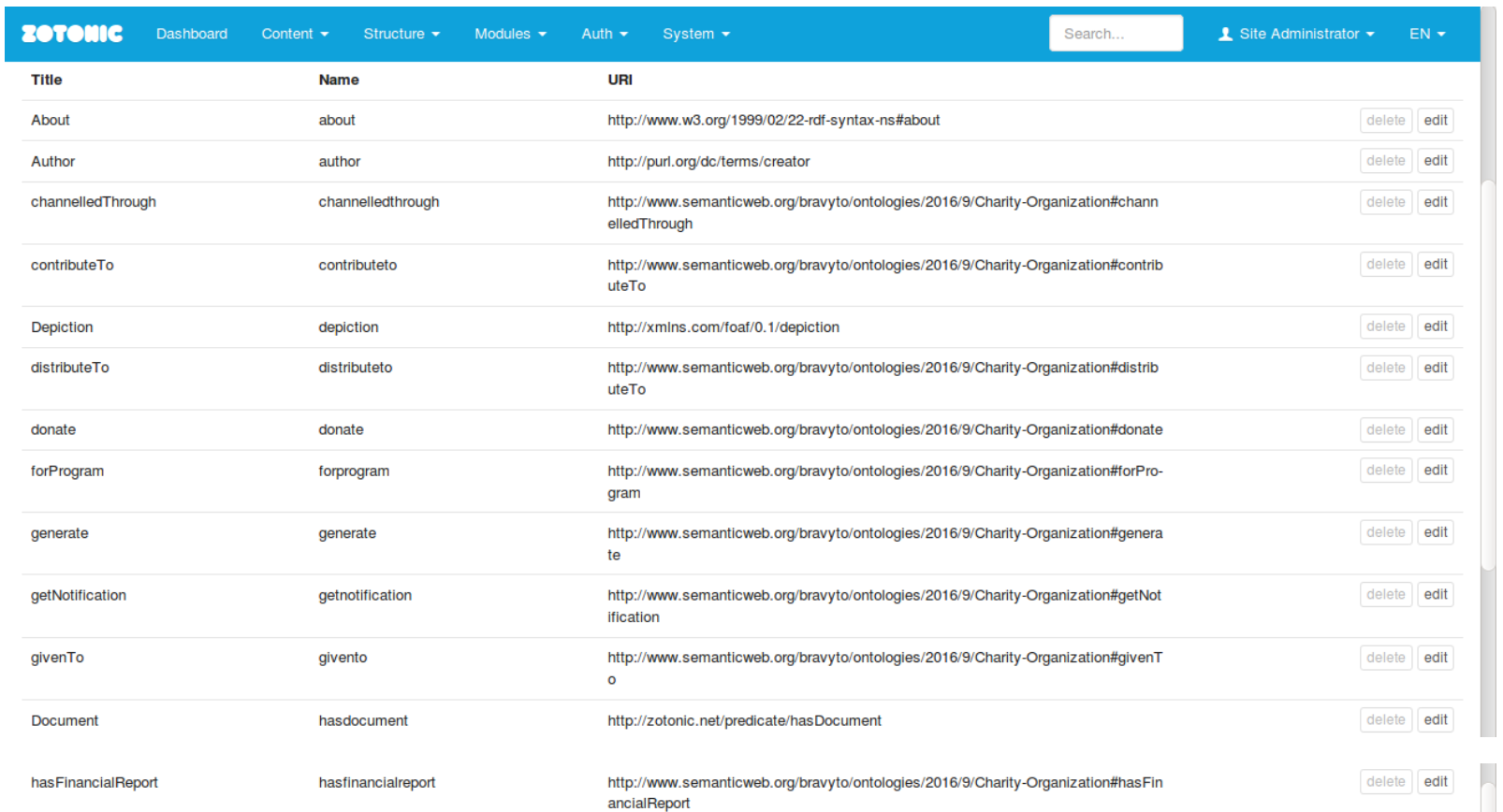}
	\caption{Screenshot of Predicate list in Zotonic page}
	\label{predicateaisco}
\end{figure}

\section{\textbf{Concluding Remark}}
In this paper, we confirmed that it is applicable to translate a feature model into a product as part of application realisation in SPL. The mechanism, firstly translates the feature diagram into OWL ontology and then the ontology is included in Zotonic to produce an application. Thus, this study also enhances our understanding that the ontology is not only used as a formal semantic representation for a feature diagram, but also for producing an application.

Considering changing scenario in a feature model (e.g add or remove mandatory features, add or remove optional features and so on), study from Dermeval et al indicated that this ontology modelling will require more time and less flexible compared to OWL individuals approach \cite{tenorio}. However, expressing feature model in an ontology language like OWL gives the advantage for the purpose of inconsistency checking over the feature classes as already conducted by \cite{verifying}. Since the features will be well maintained, reusability components or features is predominantly possible to develop others software for different stakeholders in a product line. 

In addition, such mechanism leads to the linked data with the domain is in a product line. The URI produced from each feature could be linked to our OWL and aligned to the existing ontologies.  We can generally retrieve the data from the application's web service and linked to the OWL ontology. The OWL could be aligned with existing organisation ontology such as ORG to realise linked open data.

\end{document}